# Understanding how off-stoichiometry promotes cation mixing in LiNiO$_2$

Cem Komurcuoglu[1,2], Yunhao Xiao[3], Xinhao Li[1,2], Joaquin Rodriguez-Lopez[4], Zheng Li[3], Alan C. West[1,2] and Alexander Urban[1,2,*]

[1]Department of Chemical Engineering, Columbia University, New York, NY 10027.
[2]Columbia Electrochemical Energy Center, Columbia University, New York, NY 10027.
[3]Department of Mechanical Engineering, Virginia Polytechnic Institute and State University.
[4]Department of Chemistry, University of Illinois Urbana-Champaign.

**Abstract**

Although LiNiO$_2$ is chemically similar to LiCoO$_2$ and offers a nearly identical theoretical capacity, LiNiO$_2$ and related Co-free Ni-rich cathode materials suffer from degradation during electrochemical cycling that has prevented practical use in Li-ion batteries. The observed capacity decay of LiNiO$_2$ has been attributed to the formation of structural defects via Li/Ni cation mixing that reduces cyclability and leads to poor capacity retention. Herein, we investigate the kinetics and thermodynamics of Li/Ni mixing in ideal LiNiO$_2$ and off-stoichiometric Li$_{1-z}$Ni$_{1+z}$O$_2$. We find that ideal LiNiO$_2$ is stable against cation mixing with similar characteristics as LiCoO$_2$. Li/Ni mixing is promoted by extra Ni in the Li layers that cannot be avoided in synthesis. Our study elucidates the crucial role of extra Ni atoms on Li sites in the cation mixing mechanism, an insight that can inform the development of Co-free cathode materials.

## 1. Introduction

The demand for lithium-ion batteries is projected to increase by 300% by the end of the decade.[1] Current high-energy-density lithium-ion batteries rely on cobalt-containing cathode materials, but the geographical localization of cobalt has created an urgency to shift away from its use. Nickel is chemically similar to cobalt but more abundant and its mining and supply chain is diversified over several countries and continents.[1,4] While nickel-rich materials, such as LiNi$_x$Co$_{y<x}$Al$_{1-x-y}$O$_2$ (NCA) and LiNi$_x$Mn$_y$Co$_{y<x}$O$_2$ (Ni-rich NMC), reduce the need for cobalt they do not eliminate it entirely.[2,3] Unfortunately, Co-free Ni-based cathode materials suffer from bulk and surface instabilities.[5,6]

The structural stability of Ni-rich cathode materials decreases with the fraction of Ni content.[7] In particular, Ni-rich cathode materials may undergo surface reconstructions and oxygen loss and can exhibit strong Li/Ni mixing after charge-discharge cycling.[8–11]

Given the structural complexity of layered Ni-rich cathode materials, such as NMC and NCA, modeling degradation with atomistic details is a significant challenge.[12,13] On the other hand, the Ni-rich cathode materials are isostructural to LiCoO$_2$ and LiNiO$_2$, *i.e.*, they are composed of alternating layers of Li atoms and transition metals (TM) atoms, separated by layers of oxygen atoms.[14–16] LiNiO$_2$ can be regarded as a model material for Co-free Ni-rich cathodes for understanding the underlying atomic-scale mechanisms in other Ni-rich cathode materials.

LiNiO$_2$ exhibits capacity fading as it is electrochemical cycled,[17–22] which has previously been attributed to Li/Ni-mixing, *i.e.*, the migration of Ni atoms from their initial site onto a vacant Li site within the Li layer.[6,8] Several reasons have been proposed why Li/Ni mixing may lead to capacity fading: Dahn et al. argued that the presence of Ni atoms in the Li layer (denoted Ni$_{Li}$ in the following) is detrimental to the Li diffusivity as the Ni$_{Li}$ atoms block possible Li migration pathways,[20] leading to a reduced capacity as less Li can be extracted. Kang et al. showed that the Ni$_{Li}$ atoms reduce the equilibrium height of the Li layer, *i.e.*, the Li-slab distance, affecting Li diffusion in the proximity.[23] This slab contraction can also prevent Li atoms from reintercalation into the host structure on discharge.[13] In addition, the activation energy for Ni migration in LiNiO$_2$ has been reported to lie between the values for LiCoO$_2$ and LiMnO$_2$, indicating that LNO has an inferior rate capability than LCO but should be superior to LMO.[24,25]

The origin of the observed Li/Ni-mixing remains controversial. Some studies conclude that Li/Ni mixing occurs exclusively at the surface of LiNiO$_2$ upon oxygen release.[26,27] Other studies indicate that Li/Ni mixing can also occur in the bulk of the material.[28–31] The cation-mixed spinel structure is more stable than the layered structure for the composition Li$_{0.5}$NiO$_2$,[32] but despite this thermodynamic preference, the layered-to-spinel transition proceeds slowly and only at high temperatures.[33–35]

LiNiO$_2$ made with conventional solid-state synthesis generally deviates from the ideal composition containing extra Ni on Li sites (Ni$_{Li}$).[2–5] Delmas et al. found that Li/Ni mixing is related to the degree of such off-stoichiometry but investigated only a narrow range of compositions without proposing an underlying mechanism.[36] To make sense of these seemingly contradictory observations, we employed first-principles calculations to determine the energetics and kinetics of Li/Ni mixing in bulk LiNiO$_2$. We considered the impact of the Li-concentration, Ni valence states, and Ni$_{Li}$ off-stoichiometries. Based on the mechanism we deduced from first principles, we carried out basic Monte–Carlo simulations of Li/Ni mixing during cycling for comparison with experiments in the future.

## 2. Methods

### 2.1. Density-functional theory calculations

All density-functional theory (DFT)[6,7] calculations were performed within the projector-augmented-wave (PAW)[8] method as implemented in the Vienna *ab initio* Software Package (VASP).[9–12] Unless otherwise indicated, the exchange-correlation functional by Perdew, Burke, and Ernzerhof (PBE)[13] was used with an additional rotationally invariant Hubbard-$U$ correction[14,15] of $U = 6$ eV to counteract the self-interaction error in the Ni $d$ states.[16,17] For benchmark purposes, we also performed calculations with other $U$ values ($U = 0$ eV and $U = 5$ eV) and with the regularized strongly constrained and appropriately normed (r$^2$SCAN) meta-GGA functional[18,19] without $U$ correction and with U = 0.41 eV[20] for the Ni $d$ states. These benchmarks are described in section 3.3. The plane-wave energy cutoff was 520 eV, and Brillouin zone sampling used a Γ-centered $k$-point mesh with $N_i$ subdivisions along the reciprocal lattice vector $\boldsymbol{b}_i$, with $N_i = \lfloor \max(1, R_k|\boldsymbol{b}_i| + 0.5)) \rfloor$ and $R_k = 25$ Å. The electronic convergence criterion was $10^{-5}$ eV, and the criterion for structure optimizations was $10^{-4}$ eV Å$^{-1}$ for the atomic forces.

The *enumlib* library was used for structure enumeration.[21–23] Structure enumeration was used to search for possible Li orderings and arrangement of defects. More details are given in section 2.3. Nickel oxidation states were assigned through spin integration.[24] Structure analysis and manipulation were carried out with the Atomic Simulation Environment (ASE)[25] and the Python Materials Genomics (PyMatgen) package.[26]

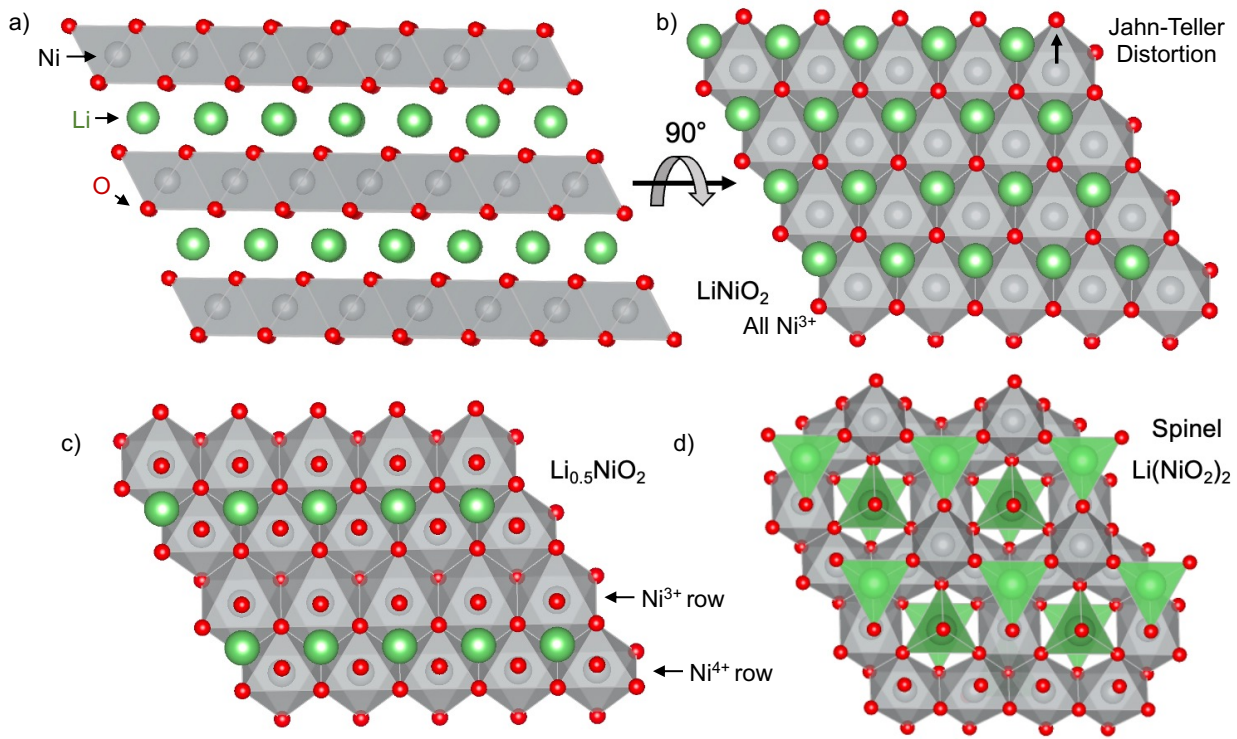

**Figure 1**: Schematic of LiNiO$_2$ showing Ni, Li, and O layers in a) a side view and b) in a projection perpendicular to the layers. The direction of the Jahn–Teller distortions is indicated by an arrow. c) Visualization of the Li/vacancy ordering in Li$_{0.5}$NiO$_2$. Upon removal of 50% of Li in LiNiO$_2$, 50% of Ni$^{3+}$ atoms are oxidized to Ni$^{4+}$. Ni$^{3+}$ and Ni$^{4+}$ are arranged in rows. d) Schematic of the Li(NiO$_2$)$_2$ spinel structure, in which 25% of the Ni atoms have migrated to the Li layers and the Li atoms are in tetrahedral sites.

## 2.2. Atomic structure

LiNiO$_2$ forms in a layered α-NaFeO$_2$-type structure (Error! Reference source not found.**a**). Ideal layered LiNiO$_2$ consists of sheets of edge-sharing NiO$_6$ octahedra with an ABC stacking sequence between which Li atoms are inserted that can be reversibly deintercalated.[27–30] Macroscopically, the space group of the LiNiO$_2$ structure is $R\bar{3}m$, implying that both Ni and Li sites are perfectly octahedral. However, computational studies strongly suggest that individual Ni sites in LiNiO$_2$ exhibit Jahn–Teller distortions, and the apparent higher symmetry seen on the macroscopic scale is due to temperature fluctuations and disorder of the distortions.[31–33] The precise ground-state structure of LiNiO$_2$ remains controversial, also because computational predictions from electronic-structure theory are to some extent dependent on methodological details.[34]

For the present study, however, the fully lithiated LiNiO$_2$ is irrelevant. Li/Ni mixing requires vacant Li sites onto which Ni can migrate. When Li is extracted from LiNiO$_2$, Ni$^{3+}$ is oxidized to Ni$^{4+}$, and spectroscopic studies[4,35] show no evidence for Ni$^{2+}$, *i.e.*, 2Ni$^{3+}$ → Ni$^{2+}$ + Ni$^{4+}$ disproportionation does not occur in significant quantity. We, therefore, focus here on Li$_{1-x}$NiO$_2$ derived from a LiNiO$_2$ structure with the C2/m space group shown in **Figure 1b,** which exhibits a collinear ordering of Jahn–Teller distorted Ni$^{3+}$ sites. **Figure 1c** shows Li$_{0.5}$NiO$_2$, one of the delithiation products of LiNiO$_2$ on which we focused.

Off-stoichiometric LiNiO$_2$ has the same general O and Ni framework as the ideal layered structure.[36] Here, we consider extra Ni in off-stoichiometric Li$_{1-z}$Ni$_{1+z}$O$_2$, in which Li atoms on the Li 3b sites in the Li layer were replaced by Ni atoms, creating Ni$_{Li}$ anti-site defects.

At the composition Li$_{0.5}$NiO$_2$, the spinel structure (**Figure 1d**) is the thermodynamic ground state, and the layered structure is only metastable.[27,37–39] Spinel has the same oxygen framework as the layered structure, but the Li atoms occupy 1/8 of the tetrahedral sites instead of octahedral sites, and 25% of the Ni atoms are in what would be the Li layers in the layered structure.[40,24,37] The ideal spinel structure has cubic symmetry with the space group $Fd\bar{3}m$, although the symmetry can be reduced when Jahn–Teller distortions are present.[40,41]

Each octahedral Ni site in the Ni layer of LiNiO$_2$ is face-sharing with two tetrahedral sites in the two adjacent Li layers. Previous extensive computational work determined that Ni migration from the Ni layer to the Li layer proceeds via the formation of Li–Ni *dumbbells*, in which the Ni atom migrates into one of the neighboring tetrahedral sites, and a Li migrates into the other.[24,27,42] To form the structural motif of the spinel structure, the Ni atom in the tetrahedral site is replaced by a Li atom and migrates further to a neighboring octahedral site. This mechanism is summarized in **Figure 2** .

### 2.3. Migration pathway calculations

To analyze the MEP for Ni migration in $Li_{0.5}NiO_2$ we compared different migration pathways by calculating the formation energies of intermediates, $E = E_{\text{intermediate}} - E_{\text{layered,pristine}}$, ruling out high energy intermediates and selecting the candidates with the lowest formation energies. We searched for possible Ni-migration pathways in layered $Li_{0.5}NiO_2$ by manually constructing intermediates. We investigated all the migration mechanisms in $\alpha$-$NaFeO_2$-type layered materials proposed by Reed et al.[43]

For the energetically most plausible Ni migration pathways (**Figure 2**), we performed nudged elastic band (NEB) calculations (see details below) to determine transition states and activation energies. We calculated the energetics of the MEP considering the oxidation state of the migrating Ni, considering the local environment of the migrating Ni with respect to Li, i.e., the number of Li vacancies nearby. Additionally, we investigated the energetics of Ni migration with respect to the Li concentration, comparing $Li_{0.5}NiO_2$ with $Li_{0.25}NiO_2$ and $Li_{0.125}NiO_2$. To investigate the formation of multiple defects, we accounted for possible site combinations manually for the formation of two spinel defects and via systematic enumeration with the *enumlib* package for greater numbers of defects.[21–23]

To investigate the $Li_{0.5}NiO_2$ layered to spinel $Li(NiO_2)_2$ phase transformation, we calculated the formation energy of each symmetrically distinct arrangement of defects (spinel motifs) by systematic enumeration. This allowed us to investigate the stepwise formation of the spinel starting from the layered structure. The energy of each step in the transformation is

$$E = \frac{E_{\text{Defect(x\%)}} - E_{\text{Layered}}}{N_{\text{fu}}}$$

where $E_{\text{Layered}}$ is the DFT energy of the ideal layered structure and $E_{\text{Defect(x\%)}}$ represents the energy of a structure with a defect concentration of x%, and $N_{\text{fu}}$ is the number of $Li_{0.5}NiO_2$ formula units.

To investigate the impact of extra Ni in the Li layer on Ni migration, we constructed an off-stoichiometric $Li_{0.5-z}Ni_{1+z}O_2$ structure, by introducing one Ni atom in the Li-layer, *i.e.*, creating a $Ni_{Li}$ defect, accounting for the Li-blocking effect around the $Ni_{Li}$ found by Sicolo et al.[44] We then investigated Ni migration around the $Ni_{Li}$ defect by constructing the Li/Ni dumbbell and spinel-defect intermediates found for ideal $Li_{0.5}NiO_2$. This was repeated for several Ni atoms with different oxidation states in the proximity of the $Ni_{Li}$.

For all calculations, a supercell with 32 $Li_{0.5}NiO_2$ formula units was employed (a 4x4x2 supercell of a primitive $LiNiO_2$ unit cell). NEB [45–48] and artificial-intelligence-driven (AID) NEB[49] calculations were performed using a smaller supercell with 8 $Li_{0.5}NiO_2$ formula units, but the formation energy of all activated states was recalculated using the original larger (4x4x2) cell. The climbing-image NEB method[45–48] was used. We used the NEB and AID-NEB implementations in the Atomic Simulation Environment (ASE).[25] The number of images in the NEB calculations ranged from 4 to 7, and calculations were considered converged when the absolute value of the force of all individual atoms was $f_{max} < 0.05$ eV/Å. The *hop* of an individual atom was generally considered a local event, and lattice parameters were kept constant except where indicated otherwise.

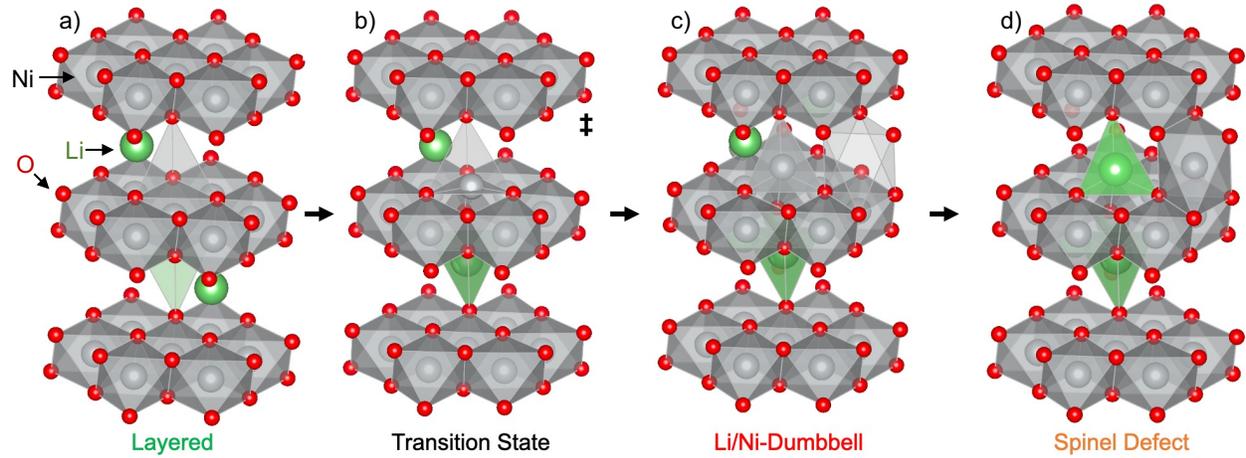

**Figure 2**. Mechanism for Ni migration from the Ni to the Li layer in $Li_{1-x}NiO_2$ as determined in previous work.[24,27,42] a) In the layered structure, each Ni and Li atom occupies an octahedral site. b) Ni migration to the Li layer proceeds via a triangular transition state in which the Ni atom is coordinated by three oxygen atoms. c) This leads to the formation of Li–Ni dumbbells with Li and Ni atoms in tetrahedral sites. d) When the tetrahedral Ni is replaced by a Li atom and migrates further onto an octahedral site, the local motif of the spinel structure, *i.e.*, a spinel defect in the layered structure, is obtained.

## 3. Results

### 3.1. Ni migration in ideal LiNiO$_2$

As reviewed above, Ni migration in ideal LiNiO$_2$ has been previously investigated with DFT.[24,27,50,51] However, these older studies were limited to comparably small cell sizes and hand-picked structures because of the limited computational power at the time. To achieve a consistent standard that the results for off-stoichiometric $Li_{1-z}Ni_{1+z}O_2$ can be compared with, we determine here (i) the defect formation energies of isolated Li–Ni dumbbell and spinel defects, (ii) the impact of the spinel defect concentration on the Ni migration energetics, and (iii) the impact of the Li concentration.

#### 3.1.1 Dumbbell and spinel defect formation in $Li_{0.5}NiO_2$

**Figure 3a** shows the defect formation energies of isolated Li–Ni dumbbells (0.70 eV) and spinel defects (0.62 eV) in $Li_{0.5}NiO_2$. Both energies are substantially positive, and the spinel defect is more stable than the dumbbell. Also shown in the figure is the energy of the transition state for the dumbbell formation, which is 1.49 eV above the initial layered structure. See **Figure S1** for the MEP from the corresponding NEB calculation. Together, the high defect formation and transition-state energies imply that isolated dumbbell and spinel defects occur in exceedingly low concentrations. Ignoring entropic effects and interactions between defects, a low spinel defect concentration of approximately $\exp(-0.62 \text{ eV}/k_BT) \approx 10^{-11}$ would be expected at room temperature.

However, as discussed above, for the composition $Li_{0.5}NiO_2$ the spinel structure is energetically favored over the layered structure, and the DFT predicts it to be –0.11 eV per formula unit more stable. As seen in **Figure 3b**, the average defect formation energy indeed decreases with an increasing spinel defect concentration, showing that spinel defects stabilize each other. Defining the spinel defect concentration such that 0% corresponds to the layered structure and 100% to the spinel structure, the mean spinel defect formation energy changes from 35 meV/f.u. for a defect concentration of 25% to 8 meV/f.u. for 50% and −9 meV/f.u. for 75%. Hence, a defect concentration of more than 50% is required for the formation of additional defects to become energetically downhill.

In summary, the spinel defect formation energetics indicate that the layered-to-spinel phase transition of $Li_{0.5}NiO_2$ is a highly concerted process with a large activation energy that cannot explain partial Li/Ni mixing. Based on our analysis, we would not expect any significant Ni migration in ideal $Li_{0.5}NiO_2$.

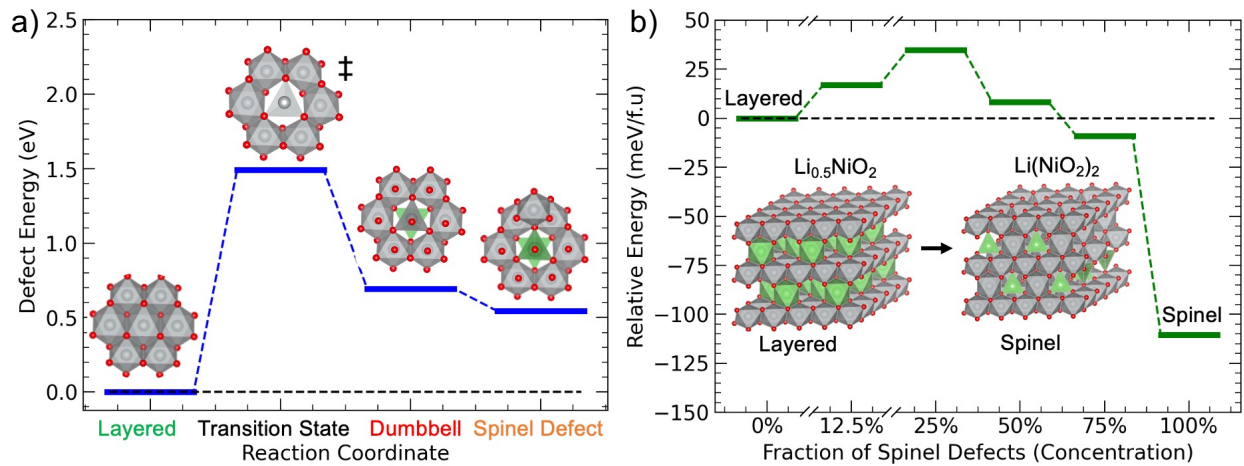

**Figure 3**. Li–Ni dumbbell and spinel defect formation in $Li_{0.5}NiO_2$. a) Defect formation energies along the Ni migration pathway of **Figure 2** in a large 4×4×2 supercell, corresponding to a defect concentration of 12.5%. The formation of isolated Li–Ni dumbbell and spinel defects is energetically uphill. b) Relative energy of layered structures with increasing concentration of spinel defects. The phase transition from the layered to the spinel structure becomes energetically downhill only after more than 50% of the layered structure has converted to spinel. All energies are listed in **Table S1**.

### 3.1.2 *Impact of the Li content and Ni oxidation state on Ni migration*

While the energy difference between the layered and spinel structure is greatest at the composition $Li_{0.5}NiO_2$[37], it is plausible that a greater Li-vacancy concentration could promote Ni migration into the Li layer. To test this hypothesis, we repeated the defect formation energy calculations of the previous section for two more compositions with lower lithium content, $Li_{0.25}NiO_2$ and $Li_{0.125}NiO_2$. The results are shown in **Figure 4a** (see **Table S1** and **Table S2** for all energies), indicating that lower Li content does not make Ni migration energetically more feasible and that the lowest defect formation energies are seen in $Li_{0.5}NiO_2$.

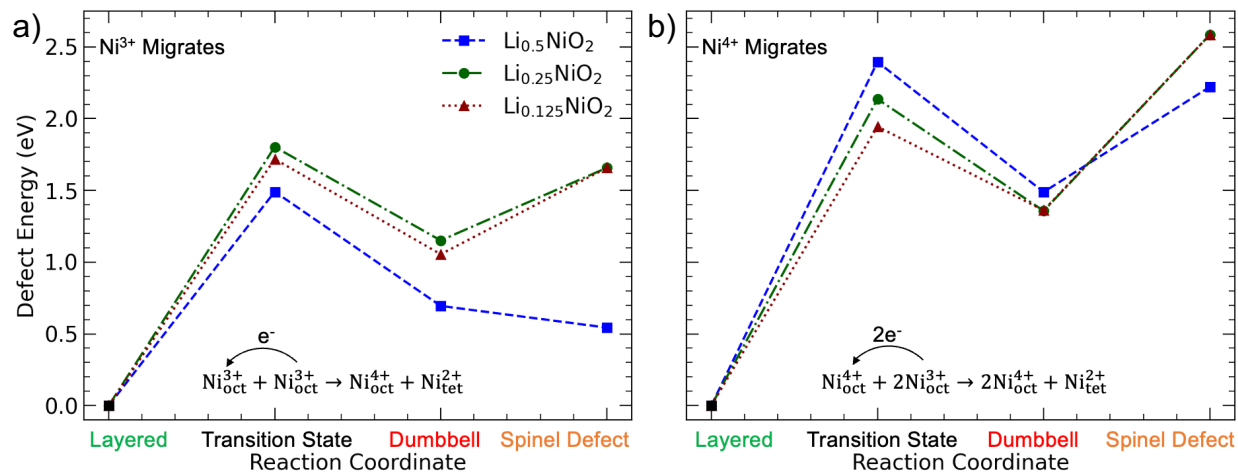

**Figure 4**: Impact of the Li content and Ni oxidation state on the formation of spinel defects and Li–Ni dumbbells. **a)** $Ni^{3+}$ migration is most facile at the spinel composition $Li_{0.5}NiO_2$. **b)** $Ni^{4+}$ migration is generally less favorable than $Ni^{3+}$ migration. All data is listed in **Table S1** and **Table S2**.

Upon Li extraction from $LiNiO_2$, Ni oxidizes, *i.e.*, the valence state of some of the Ni atoms changes from 3+ to 4+. This means two different Ni species in the material can potentially migrate to the Li layer. As shown in **Figure 4b**, $Ni^{3+}$ migration is generally energetically more favorable.

Our data, thus, indicates that Li/Ni mixing should not occur in ideal $LiNiO_2$ at any Li content. Although the layered-to-spinel phase transition is energetically favorable for $Li_{0.5}NiO_2$, it is a concerted process that cannot explain incomplete/partial Li/Ni mixing. Deeper delithiation does not change this picture, also because Ni migration becomes less feasible as more Ni is oxidized to a 4+ valence state.

### 3.2. Ni migration in off-stoichiometric $Li_{1-z}Ni_{1+z}O_2$

As reviewed in the introduction section, $LiNiO_2$ synthesized via conventional solid-state routes typically contains extra Ni so that the true composition is $Li_{1-z}Ni_{1+z}O_2$. In this section, we investigate the impact of such extra Ni on the energetics of Ni migration.

#### 3.2.1 Structure and valence states in the presence of extra Ni atoms in the Li layer

**Figure 5** shows the geometry and Ni oxidation states in a $Li_{1-z}Ni_{1+z}O_2$ structure model with an extra Ni atom on one of the Li sites in the Li layer. $Ni_{Li}$ defect results in a contraction of the Li layer near the extra Ni from an average O–O plane distance of 2.97 Å to a local distance of 2.87 Å (3.5%) (**Figure 5a**). As also previously reported[4,52,53], this slab contraction, as well as the Li–Ni electrostatic repulsion, destabilize the neighboring Li sites, which are generally vacant once the overall Li content permits it.

The $Ni_{Li}$ site is Jahn–Teller distorted, and integration of the spin-density difference around the extra Ni shows no unpaired electrons. This indicates that the extra Ni atom is in a 2+ valence state, in agreement with intuition based on the ionic radius and electrostatic arguments as discussed in Section 3.1.2. The extra Ni atom also affects the Ni atoms in the neighboring Ni layers: even when the structure is delithiated to the overall composition $Li_{0.4242}Ni_{1.0315}O_2$, one of the corner-sharing Ni atoms remains in a 2+ oxidation state, and the Ni atoms are no longer equivalent (**Figure 5b**). The Ni–O bond lengths and results from spin integration of the extra Ni atom and all Ni atoms highlighted in (**Figure 5b**) are given in and **Figure S2**.

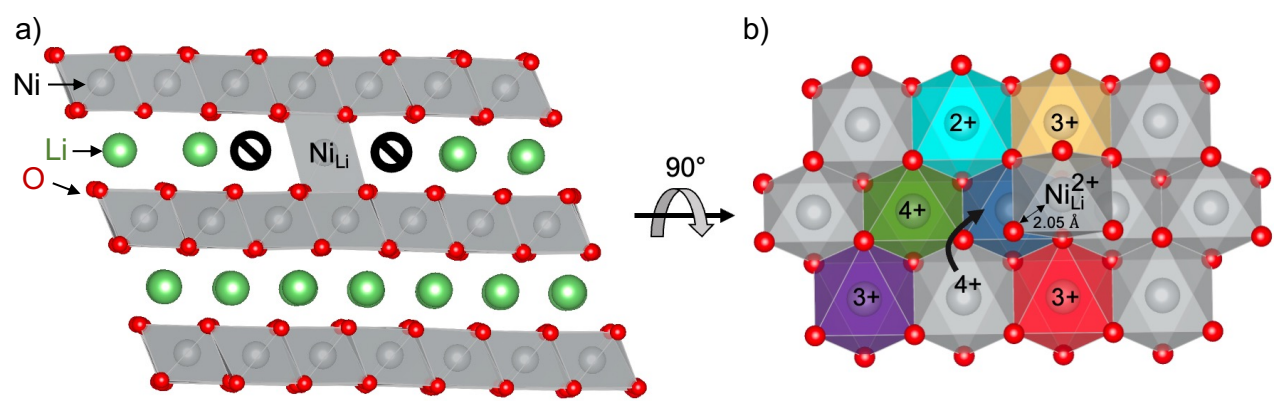

**Figure 5.** Extra Ni atom on a Li site within the Li layer ($Ni_{Li}$) in off-stoichiometric $Li_{1-z}Ni_{1+z}O_2$. **a)** The extra Ni atom leads to a contraction of the Li slab. Once sufficient Li has been extracted, the Li sites neighboring the extra Ni are vacant. **b)** The $Ni_{Li}$ defect leads to the reduction of a corner-sharing Ni atom to the 2+ valence state in one of the neighboring Ni layers. Upon delithiation, this $Ni^{2+}$ does not immediately oxidize even when nearby Ni atoms are oxidized from 3+ to 4+. The color coding in the figure distinguishes the different Ni species near a $Ni_{Li}$ defect in a $Li_{0.4242}Ni_{1.0315}O_2$.

### 3.2.2 *Ni migration in the presence of extra Ni*

To determine the impact of a $Ni_{Li}$ defect on Li/Ni mixing, we calculated the Li–Ni dumbbell and spinel-defect energies of the different Ni species in the structure of **Figure 5b**. In our calculations, we generally migrated the Ni atoms to a second-nearest neighbor site of the $Ni_{Li}$ atom to avoid the electrostatic repulsion between neighboring Ni atoms in the Li layer. The results are visualized in **Figure 6** and listed in **Table S3**.

As seen in the figure, the defect formation energies vary widely between the different Ni atoms. The most obvious trend is an increasing defect formation energy with an increasing Ni oxidation state.

Remarkably, the dumbbell and spinel defect formation energies are negative (–0.09 eV and –0.04 eV, respectively) for the $Ni^{2+}$ atom in a corner-sharing site with the $Ni_{Li}$ atom. For the corner-sharing $Ni^{3+}$ atom, the dumbbell formation energy is also 44% lower (0.39 eV) than in ideal $Li_{0.5}NiO_2$ (0.69 eV), and the spinel defect formation energy is approximately 0.00 eV, while it was 0.62 eV without extra Ni. This means, $Ni^{2+}$ migration to the Li layer is thermodynamically favorable, and $Ni^{3+}$ migration is a slow process at room temperature due to the sizable activation energy, but in equilibrium, one would expect the corner-sharing $Ni^{3+}$ to be evenly distributed across the Ni and Li layers.

The formation energies of multiple spinel defects for stoichiometric $Li_{0.5}NiO_2$ and $Li_{0.4242}Ni_{1.03125}O_2$ are shown in **Figure S3**. The formation of two spinel defects in $Li_{0.4242}Ni_{1.03125}O_2$ requires a net positive energy of (0.27 eV), which is, however, still lower than the formation of a single spinel defect in stoichiometric $Li_{0.5}NiO_2$ (0.54 eV). The formation of three spinel defects in the presence of $Ni_{Li}$ defect (0.63 eV) is energetically comparable to the formation of a single spinel defect in $Li_{0.5}NiO_2$ (0.54 eV).

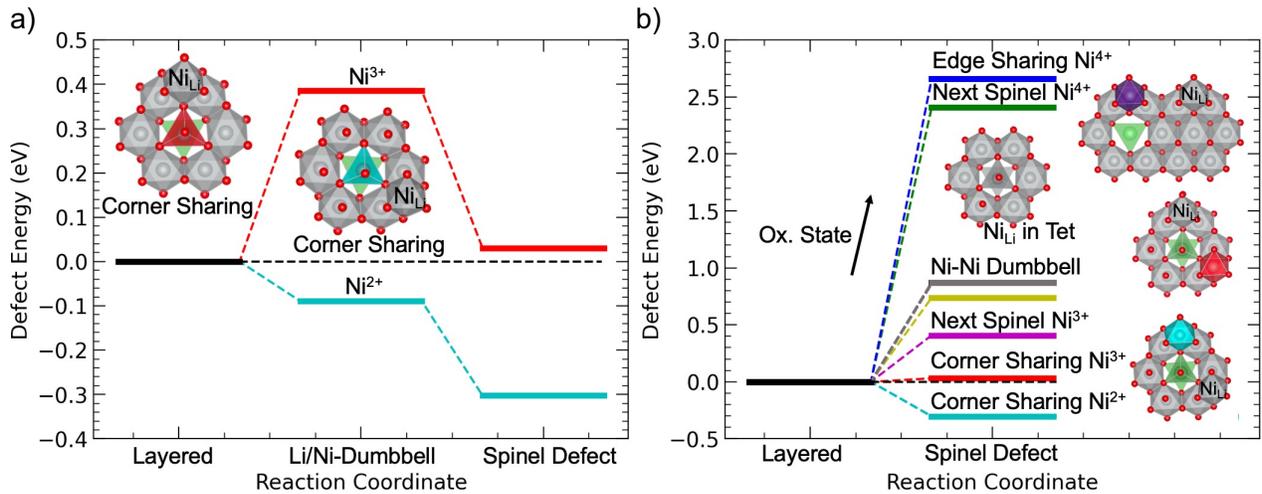

**Figure 6.** Li–Ni dumbbell and spinel-defect formation energies for migration of the various Ni species of **Figure 5b** in off-stoichiometric $Li_{0.4242}Ni_{1.03125}O_2$. **a)** Dumbbell and spinel defect formation energies in the presence of a $Ni_{Li}$ defect for migrating $Ni^{2+}$ and a $Ni^{3+}$ cations that are corner-sharing with the $Ni_{Li}$ defect. **b)** Spinel defect formation energies for the migration of various Ni species in the presence of a $Ni_{Li}$ defect.

### 3.3. Robustness of the DFT approach

$LiNiO_2$ and its derivatives are challenging materials for DFT to model. Due to the intrinsic self-interaction error, uncorrected semi-local DFT fails to capture the localization of the Ni 3d electrons on the metal centers, and the calculation results reported above employed a Hubbard-$U$ correction term to compensate for this effect. However, recent work has highlighted that even small differences in the employed $U$ value in PBE+$U$ calculations can lead to qualitatively different results for some properties of $LiNiO_2$.[34] Therefore, we performed a sensitivity analysis to determine how the defect formation energies vary with the $U$ value to increase our confidence in the computational predictions. In addition, we repeated some of the key calculations with the r²SCAN meta-GGA functional, which is often more accurate than PBE for the properties of transition-metal oxides.[54–57] The computational details are described in the methods section.

All results reported above were obtained from PBE+$U$ calculations with $U$(Ni 3d) = 6 eV, the $U$ value recommended in reference 58. On the other hand, in a substantial body of literature, a $U$ value of 5 eV is used for the Ni 3d states. **Figure S4a** compares spinel defect formation energies in ideal $Li_{0.5}NiO_2$ calculated with PBE+$U$ ($U$ = 5 eV) and r²SCAN+$U$ ($U$ = 0.41 eV [55]) with the values of Section 3.1. While the absolute value of the energies differ by up to 0.24 eV, the general trend is consistent across the three methods, showing substantial, positive defect formation energies in ideal $Li_{1-x}NiO_2$. **Figure S4b** shows an equivalent analysis for off-stoichiometric $Li_{0.4242}Ni_{1.03125}O_2$. Here, PBE+$U$ ($U$ = 5 eV) predicts a slightly positive energy for $Ni^{2+}$ spinel defects (0.08 eV), whereas the other methods predict a negative defect formation energy. Note that r²SCAN+$U$ predicts a significantly lower energy than PBE+$U$ ($U$ = 5 eV).

Taken together, we conclude that all three DFT approaches predict high defect formation energies in ideal $LiNiO_2$, confirming the conclusion of Section 3.1 that partial Li/Ni mixing in ideal $LiNiO_2$ is thermodynamically unfavorable. The three methods also agree that extra Ni in the Li layer creates $Ni^{2+}$ atoms and facilitates the migration of these into the Li layer. The extent to which this process is thermodynamically favored cannot be conclusively decided.

### 3.4. Simulation of Li/Ni mixing in $Li_{1-z}Ni_{1+z}O_2$ during the first charge

Based on the predicted energy profiles for Ni migration in off-stoichiometric $Li_{1-z}Ni_{1+z}O_2$, we make the following observations:

1. The activation energy for Ni migration increases with the oxidation state. $Ni^{2+}$ migration is most facile.
2. Each extra Ni atom in a Li layer results in one $Ni^{2+}$ atom in a neighboring Ni layer.
3. The migration of this $Ni^{2+}$ atom from the Ni layer to a site in the Li layer surrounded by vacancies is thermodynamically favorable.

We can use these rules to model the increase of Ni content Li layers during the first charge of $Li_{1-z}Ni_{1+z}O_2$ if we make the following additional assumptions: (1) As synthesized, we assume that all octahedral sites are occupied, *i.e.*, we assume that the composition can indeed be written as $Li_{1-z}Ni_{1+z}O_2$. (2) We assume that Li can freely diffuse on a time scale much shorter than Ni migration, so we can assume Li vacancies will be randomly distributed.

(1) and (2) are approximations because as-synthesized $LiNiO_2$ might also exhibit vacancies, and Li atoms, extra Ni atoms, and vacancies interact during delithiation. Furthermore, the above model does not account for $Ni^{3+}$ migration to the Li layer, even though two of the benchmarked DFT approaches indicate that some amount of $Ni^{3+}$ migration, indeed between 0.5 to 1.0 atom per extra Ni, has to be expected. The model, thus, provides a *lower bound* underestimating the actual degree of Li/Ni mixing.

**Figure 7** shows the Ni content in the Li layers during the first charge as predicted by a Monte-Carlo simulation of a Li layer in off-stoichiometric $Li_{1-z}Ni_{1+z}O_2$ with $z = 0.01$, 0.05, and 0.10 using the model described above and a two-dimensional lattice with 20×20 = 400 sites that achieved converged results. As seen in the figure, the fraction of Ni in the Li layer remains unchanged during the initial delithiation since the Li vacancies are distributed such that the concentration of favorable environments for $Ni_{Li}$ defects is low. Once around 0.5 Li has been extracted, significant Ni migration is observed, the degree of which increases with the initial $Ni_{Li}$ concentration. This simulation assumed random delithiation, i.e., Li/vacancy ordering was not considered, and it, therefore, likely underestimates the Li/Ni mixing. However, we expect that it predicts realistic trends and provides a lower bound for cation mixing.

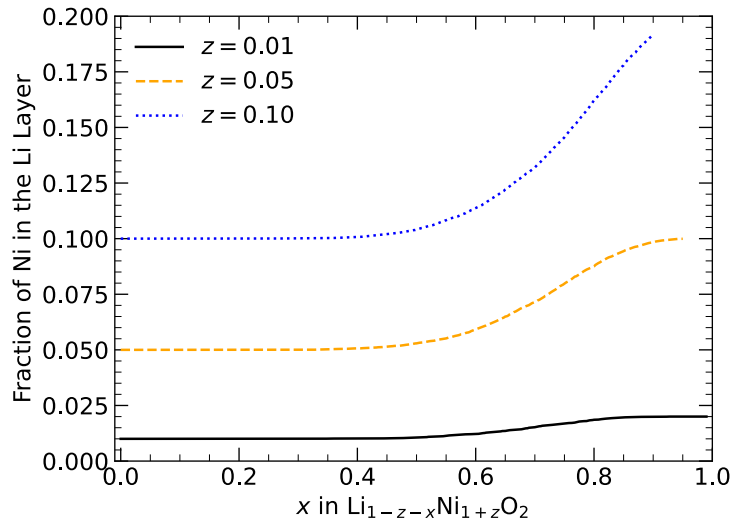

**Figure 7.** Change of the average concentration of Ni in the Li layers of $Li_{1-z}Ni_{1+z}O_2$ with different degrees of off-stoichiometry $z$ during the first charge as predicted by a lattice Monte–Carlo simulation. The Monte–Carlo simulation results are the average of 500 delithiation simulations using a two-dimensional lattice with 400 sites confirmed sufficient for size convergence.

## 4. Discussion

In the present study, we investigated how extra Ni atoms on the Li layer of $LiNiO_2$ affect the energetics of Ni migration and, thereby, Li/Ni mixing during cycling. First-principles calculations show conclusively that partial Li/Ni mixing is not thermodynamically favorable in ideal $LiNiO_2$. In contrast, it is promoted by extra Ni in $Li_{1-z}Ni_{1+z}O_2$ via two factors: (1) the extra Ni leads to the reduction of a Ni atom in a neighboring Ni layer to 2+ valence state, which can migrate to the Li layer once sufficient vacancies have been created during delithiation. (2) The Ni in the Li layer gives rise to a local contraction of the Li slab because of its smaller ionic radius compared to Li⁺ ions, which stabilizes the sites in the Li layer for further migrating Ni atoms. This mechanism is visualized schematically in **Figure 8**.

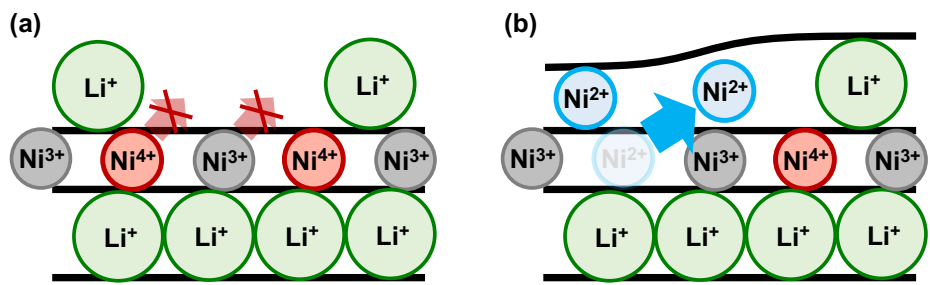

**Figure 8.** Schematic of the Ni migration mechanism identified in the present study. (a) In ideal LiNiO2, Li/Ni mixing in the bulk is unlikely because the migration of $Ni^{3+}$ and $Ni^{4+}$ from the Ni to the Li layer is thermodynamically unfavorable and subject to high activation energies. (b) In off-stoichiometric $Li_{1-z}Ni_{1+z}O_2$, extra Ni in the Li layer leads to $Ni^{2+}$ in the Ni layer, the migration of which is facile. In addition, the Ni in the Li layer causes a local Li slab contraction that stabilizes the migrating Ni atom and makes this process thermodynamically favorable.

The result that the activation energy for Ni migration increases with the Ni valence state is consistent with intuition and prior work. Reed and Ceder previously showed with electronic-structure arguments that only $Ni^{2+}$ is stable in tetrahedral sites.[24,43] For forming a Li–Ni dumbbell, $Ni^{3+}$ and $Ni^{4+}$ have to accept electrons from neighboring Ni atoms. $Ni^{3+}$ tends to undergo disproportionation, $2Ni^{3+} \rightarrow Ni^{2+} + Ni^{4+}$, but $Ni^{4+}$ is generally more stable. It would require two neighboring $Ni^{3+}$ atoms to reduce to 2+ ($Ni^{4+} + 2Ni^{3+} \rightarrow Ni^{2+} + 2Ni^{4+}$). Electron transfer from a neighboring $Ni^{3+}$ to a migrating $Ni^{3+}$ is hindered because of the substantial decrease in the ligand field stabilization energy arising from the redistribution of the electronic states ($2t_{2g}^6 e_g^1 \rightarrow t_{2g}^6 + e^4 t_2^4$). This effect is also referred to as electronic frustration.[59,60] Since Ni cannot be oxidized beyond the +4 oxidation state, $Ni^{4+}$ cannot provide additional electrons. This relationship makes Ni migration via disproportionation less likely the more Li is extracted and the greater the average Ni oxidation state becomes. Furthermore, each octahedral site in the Li layer shares six edges with octahedral Ni sites in the neighboring Ni layers. Replacing Li⁺ with Ni increases the amount of electrostatic repulsion that the Ni atoms in the adjacent Ni layers experience so that the electrostatic energy of Ni atoms in the Li layer increases with the oxidation state. Finally, the ionic radius of $Ni^{2+}$ (0.69 Å) is closer to that of Li⁺ ions (0.76 Å) than those of $Ni^{3+}$ (0.56 Å) or $Ni^{4+}$ (0.48 Å)[61], so that $Ni^{2+}$ creates less strain in the Li layer than the other Ni species.

Some early computational work predicted the formation of spinel defects, especially Li–Ni dumbbells, in ideal LiNiO2 to be energetically more favorable.[27] These studies were based on much smaller cell sizes corresponding to high defect concentrations, which agrees with our results (**Figure 3**). Recent computational work concluded that Li/Ni mixing in ideal LiNiO$_2$ would likely only appear in the surface regions upon oxygen release.[62]

Our findings for ideal LiNiO$_2$ also agree with recent studies that reported characterization results for nearly stoichiometric LiNiO$_2$.

Using *in situ* X-ray diffraction, Ikeda et al. showed that in nearly stoichiometric LiNiO$_2$, Ni migration is a reversible process in which, upon charge, in NiO$_2$, Ni atoms migrate from octahedral to adjacent tetrahedral sites.[63] During discharge, the authors report Ni atoms migrating back to their original octahedral sites, though the reversibility of Ni migration is apparently lost with repeated cycling, with some Ni atoms remaining in tetrahedral sites.

With first-principles-based Monte Carlo simulations, Xiao et al. studied the kinetics of the layered-to-spinel transformation in Li$_{0.5}$NiO$_2$ in the bulk.[64] The authors found that elevated temperatures were necessary for a significant fraction of the Ni atoms to migrate. At room temperature (300K), the authors predicted the half-life of the layered structure to be 6000h (= 250 days). These results corroborate our findings that Ni migration in ideal Li$_{0.5}$NiO$_2$ is a slow process and confirm that a significant fraction of the Ni atoms must migrate for spinel formation to occur.

One explanation for the small (but non-zero) amount of Li/Ni mixing seen in nearly stoichiometric LiNiO$_2$ could be the local disproportionation of Ni$^{3+}$ into 2+ and 4+. In the P2/c space group, half-delithiated Li$_{0.5}$NiO$_2$ contains 25% Ni$^{2+}$. Without extra Ni in the Li layer, the formation of spinel defects is energetically unfavorable even when Ni$^{2+}$ is present. However, the stability of Li–Ni dumbbells depends less on the Li slab height, so we would expect them to form. This would explain why Ni migration in highly stoichiometric LiNiO$_2$ is reversible,[63] since Ni atoms in dumbbells can readily migrate back to their original site in the Ni layer. Note, however, that the P2/c structure is destabilized by temperature effects because of its lower entropy, and no evidence for significant concentrations of Ni$^{2+}$ in the bulk of LiNiO$_2$ has been observed in spectroscopic studies to the authors' knowledge.

A seemingly counterintuitive result is that Ni$^{2+}$ in the Ni layer, near an extra Ni atom in the Li layer, does not immediately oxidize upon Li extraction. However, it has previously been argued that Ni$^{2+}$ can be stabilized by the strong superexchange interaction between two Ni atoms with opposite atomic magnetization along a 180° Ni–O–Ni motif.[65] Such superexchange interactions between the extra Ni atom in the Li layer and one of the nearby Ni atoms in a Ni layer might explain the stability of pairs of Ni$^{2+}$ ions in the Li and Ni layers during delithiation. Indeed, we find that the Ni$_{Li}$ atom prefers a magnetization opposite to that of the neighboring Ni layer (**Figure S4b**), facilitating superexchange interaction.

Our simulations focused on off-stoichiometries in the Li layer only, though as-synthesized Li$_{1-z}$Ni$_{1+z}$O$_2$ typically already exhibits an initial degree of cation mixing. Such cation mixing leads to a contraction of the Li slab in the pristine material but does not affect the average oxidation states of the Ni atoms. Extra Ni due to off-stoichiometries is needed to provide the mobile Ni$^{2+}$ species that give rise to additional Li/Ni mixing during cycling. However, future experimental studies could help further disentangle the impact of initial cation mixing and off-stoichiometry.

While we went to great length ensuring that our computational predictions are robust, a careful validation with experiments would provide an ultimate confirmation. To this end, we provided data from a Monte–Carlo delithiation simulation that can be compared with, for example, diffraction experiments. However, the Monte–Carlo algorithm we employed is basic and did not account for atomic interactions, so the results should be considered a lower bound for Li/Ni cation mixing.

## 5. Conclusions

We investigated Li/Ni cation mixing in the bulk of ideal layered LiNiO$_2$ and off-stoichiometric Li$_{1-z}$Ni$_{1+z}$O$_2$ with extra Ni in the Li layer upon delithiation using first-principles calculations. For ideal LiNiO$_2$, our results indicate that Ni migration from the Ni into the Li layer is energetically most feasible for the composition Li$_{0.5}$NiO$_2$, at which the cation-mixed spinel structure is the ground state. However, even at the spinel composition, the formation of individual Ni$_{Li}$ requires overcoming a high activation energy and is energetically uphill due to electronic frustration. Thus, partial Li/Ni mixing without a phase transition to the spinel structure is unlikely in stoichiometric LiNiO$_2$. Extra Ni in Li$_{1-z}$Ni$_{1+z}$O$_2$ changes this picture by introducing Ni$^{2+}$ in the Ni layer that can migrate into neighboring Li layers without an energy penalty. The result that extra-Ni off-stoichiometries promote Li/Ni mixing is robust with respect to the choice of computational parameters, and we provided delithiation data from simple Monte–Carlo simulations to facilitate comparison with experiments. The findings explain the previously empirically observed negative impact of off-stoichiometries and predict highly-stoichiometric single-crystalline LiNiO$_2$ to exhibit good bulk stability.

**Acknowledgments** This work was partially supported by the Alfred P. Sloan Foundation Grant No. G-2020-12650. We acknowledge computing resources from Columbia University's Shared Research Computing Facility project, which is supported by NIH Research Facility Improvement Grant No. 1G20RR030893-01, and associated funds from the New York State Empire State Development, Division of Science Technology and Innovation (NYSTAR) Contract No. C090171, both awarded April 15, 2010. We thank John. H. Harding for providing LiNiO$_2$ structures.

**Supplemental Information**

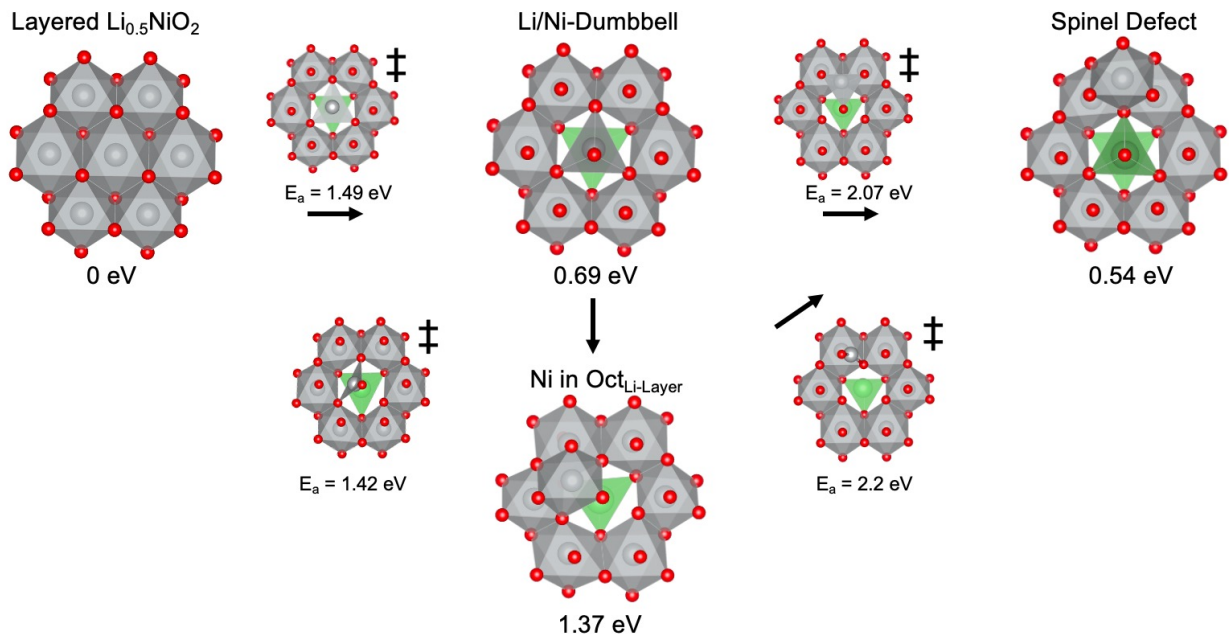

**Figure S1**: Ni-Migration pathways in $Li_{0.5}NiO_2$ and their associated activation energies ($E_a$) obtained from NEB/AID-NEB calculations. The formation energies below each motif were calculated relative to the layered $Li_{0.5}NiO_2$ structure.

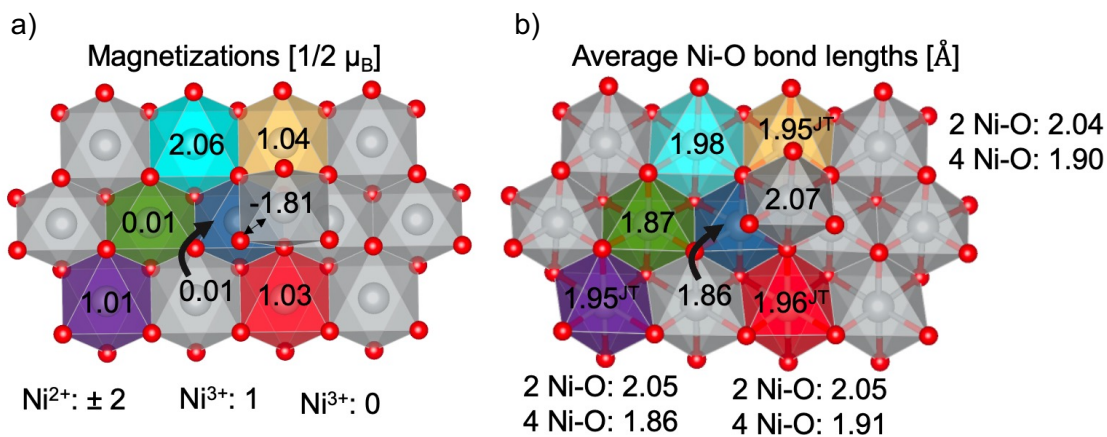

**Figure S2**: a) Atomic magnetizations of Ni atoms in off-stoichiometric $Li_{0.4242}Ni_{1.0315}O_2$ and b) the corresponding average Ni-O bond lengths. Jahn–Teller distortions are indicated with JT.

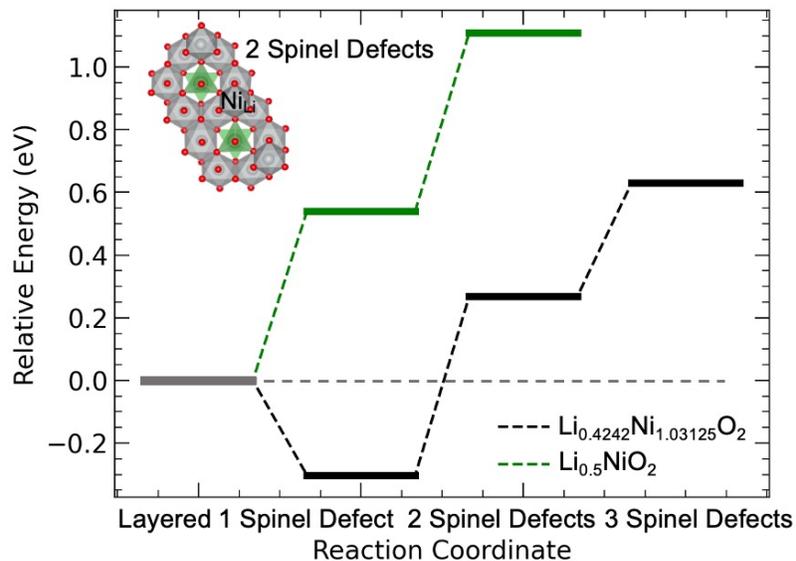

**Figure S3**: Formation energies of multiple spinel defects in $Li_{0.5}NiO_2$ and $Li_{0.4242}Ni_{1.03125}O_2$.

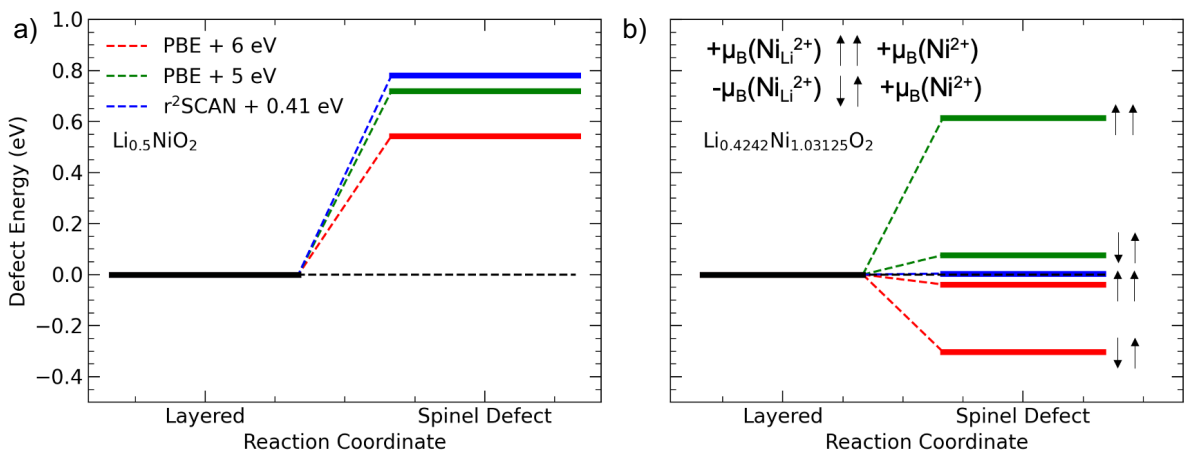

**Figure S4**: Spinel formation energies for PBE+6 eV, PBE+5 eV, and $r^2$SCAN+0.41 eV in a) stoichiometric $Li_{0.5}NiO_2$ and b) off-stoichiometric $Li_{0.4242}Ni_{1.03125}O_2$. b) The direction of the atomic magnetization $\mu_B$ (arrows) has a significant impact on the defect formation energy for Ni migration in off-stoichiometric $Li_{0.4242}Ni_{1.03125}O_2$.

**Table S1:** $Ni^{3+}$ migration: Relative energies of three key structural motifs for a migrating $Ni^{3+}$ in $Li_xNiO_2$ in eV. All energies are for PBE+U with U = 6.0 eV.

|  | $x = 0.125$ | $x = 0.25$ | $x = 0.50$ |
|---|---|---|---|
| **Transition State** | 1.718 | 1.803 | 1.491 |
| **Li-Ni Dumbbell** | 1.054 | 1.151 | 0.694 |
| **Spinel Defect** | 1.657 | 1.012 | 0.543 |

**Table S2:** $Ni^{4+}$ migration: Relative energies of three key structural motifs for a migrating $Ni^{4+}$ in $Li_xNiO_2$ in eV. All energies are for PBE+U with U = 6.0 eV.

|  | $x = 0.125$ | $x = 0.25$ | $x = 0.50$ |
|---|---|---|---|
| **Transition State** | 1.944 | 2.136 | 2.395 |
| **Li-Ni Dumbbell** | 1.361 | 1.357 | 1.488 |
| **Spinel Defect** | 2.584 | 1.183 | 2.221 |

**Table S3:** Defect Formation Energies in the presence of extra Ni.

| Defect | Relative Energy (eV) |
|---|---|
| Li/Ni-dumbbell Corner Sharing $Ni^{2+}$ | -0.089 |
| Spinel Defect Corner Sharing $Ni^{2+}$ | -0.039 |
| Li/Ni-dumbbell Corner Sharing $Ni^{3+}$ | 0.385 |
| Spinel Defect Corner Sharing $Ni^{3+}$ | 0.030 |
| Next Spinel Defect $Ni^{3+}$ | 0.409 |
| Edge Sharing Spinel Defect $Ni^{3+}$ | 0.738 |
| Ni-Ni-dumbbell | 0.871 |
| Next Spinel Defect $Ni^{4+}$ | 2.407 |
| Edge Sharing Spinel Defect $Ni4^+$ | 2.657 |